\documentclass[prb,showpacs,reprint,amsmath,amssymb,
superscriptaddress,floatfix]{revtex4-1}
\usepackage{graphicx}
\usepackage{color}
\usepackage{dcolumn}
\usepackage{bm}
\usepackage{amssymb}
\usepackage{amsmath}
\usepackage{mathrsfs}
\usepackage{braket}
\usepackage{multirow}
\usepackage{subfigure}
\usepackage{graphicx}
\usepackage{enumitem}

\definecolor{mycolor}{rgb}{0,0,0}

\def\boxeqnarray#1#2%
{{\fboxsep
0.25ex\fbox{\parbox{#1}{\begin{eqnarray}#2\end{eqnarray}}}}}

\begin{document}
\title{Josephson critical currents in annular superconductors with Pearl vortices}

\author{A. Bad\'{\i}a-Maj\'{o}s}
\email[Electronic address: ]{anabadia@unizar.es}
\affiliation{Departamento de F\'{\i}sica de la Materia
Condensada and Instituto de Ciencia de Materiales de Arag\'{o}n (ICMA), Universidad de Zaragoza--CSIC, 
Mar\'{\i}a de Luna 3, E-50018 Zaragoza, Spain}
\date{\today}

\begin{abstract}

{We investigate the influence of Pearl vortices in the vicinity of an edge-type Josephson junction for a superconducting thin-film loop in the form of an annulus, under uniform magnetic field. Specifically, we obtain the exact analytic formulation that allows to describe the circulating current density and the gauge invariant phase increment $\Delta\phi$ across the junction. The main properties of $\Delta\phi$ and their influence on the critical current pattern $I_c(B)$ are described quantitatively in terms of the loop's width to radius ratio $W/R$ and of the vortex position within the loop ${\bf r}_v$. It is shown that narrow loops ($W/R < 0.3$) may be well described by the straight geometry limit. However, such approximation fails to predict a number of distinctive features captured by our formulation, as the node lifting effect of the $I_c(B)$ pattern in wide loops or the actual influence of a vortex pinned at different positions.}

\end{abstract}
\maketitle

\section{Introduction}

The relevance of the phase of the complex order parameter in superconductivity is well known, and is receiving revived attention in the last years. To be specific, the so-called {\em flux interferometry} experiments in Josephson junctions, enabled by the new nanofabrication techniques offer a powerful probe of basic properties as well as an appealing groundwork in low temperature electronics. Thus, a number of unambiguous features related to fundamental phenomena as is the physics of vortices and their action on the nearby superconducting condensate, make clear imprints in the junctions' critical current patterns $I_c(B)$. This can be measured in magnetotransport experiments. As an example, the classical Fraunhofer-like dependence of $I_c$ on the penetrating magnetic flux\cite{ref:tinkham} undergoes strong deformations in the presence of a nearby Abrikosov vortex.\cite{ref:golodp,ref:golodn} The curve looses mirror symmetry respect to the field polarity, the central maximum may even become a minimum, and periodicity is strongly altered.

In several works,\cite{ref:clem_2011,ref:koganb,ref:koganc,ref:mironov}  it has been shown that the above features find a natural explanation when one considers the full gauge invariant phase variation ($\Delta\phi$ in what follows) across the junction. In these papers, by resorting to the Josephson's zero voltage supercurrent $I_J~=~I_c\,{\rm sin}(\Delta\phi)$, and after evaluating $\Delta\phi$, which \textcolor{mycolor}{characterizes the superconducting condensate and} ultimately depends on the magnetic flux, the above properties have been theoretically reproduced. Such studies have focused on small planar junctions between long superconducting strips, where the phase variation around the vortex\cite{ref:pearl} only affects one of the banks of the junction. 

The case of closed superconducting loops, in which the long-range coherence of the phase could be responsible of new phenomena related to the interaction with both banks has been suggested. Yet, only semiquantitative expectations were issued in Ref.\,\onlinecite{ref:koganb}, based on the extrapolation of results for long strips. In this paper, we present an analytic formulation that is used to investigate annular superconducting loops with one junction and vortices pinned at arbitrary nearby positions.

Similar to previous theoretical studies, our methodology will also encompass to find the critical current patterns with no vortex present. It will be shown that the annular geometry produces a fine structure in $I_c (B)$ that overlaps to the above mentioned vortex-mediated distortions of the Fraunhofer-like pattern. It will be important to distinguish between this effect and other vortex-free distortions reported in recent literature, as those related to: (i) material non-uniformities and temperature gradients in long junctions,\cite{ref:krasnov,ref:neils,ref:lam} (ii) the breakdown of Josephson's sinusoidal relation,\cite{ref:kurter} or (iii) the asymmetric injection of current.\cite{ref:kogana}

The work is organized as follows. In Sec.\ref{sec:const} we cast the equations of a \textcolor{mycolor}{minimal} physical model that incorporates the role of the gauge invariant phase in the response of the annuli to an applied magnetic field, either with or without vortices present. \textcolor{mycolor}{Sec.\ref{sec:solution} presents the solution to the problem. Technically,} the use of conformal mapping completely resolves the contribution of vortices and partially the contribution of sample-scale screening currents. \textcolor{mycolor}{In-depth mathematical details of the formulation are given in an appendix}. Sec.\ref{sec:globalF} puts forward the completion of the work, by showing how to compute $\Delta\phi$ based on the above and in terms of all the relevant physical parameters: geometry of the annulus, applied magnetic field and  position of the vortices. In Sec.\ref{sec:discuss} we analyze the peculiarities of the solutions for annular loops and comment on the scope of our results.

\section{Constitutive equations}
\label{sec:const}
\begin{figure}[t]
\centering
{\includegraphics[width=.5\textwidth]{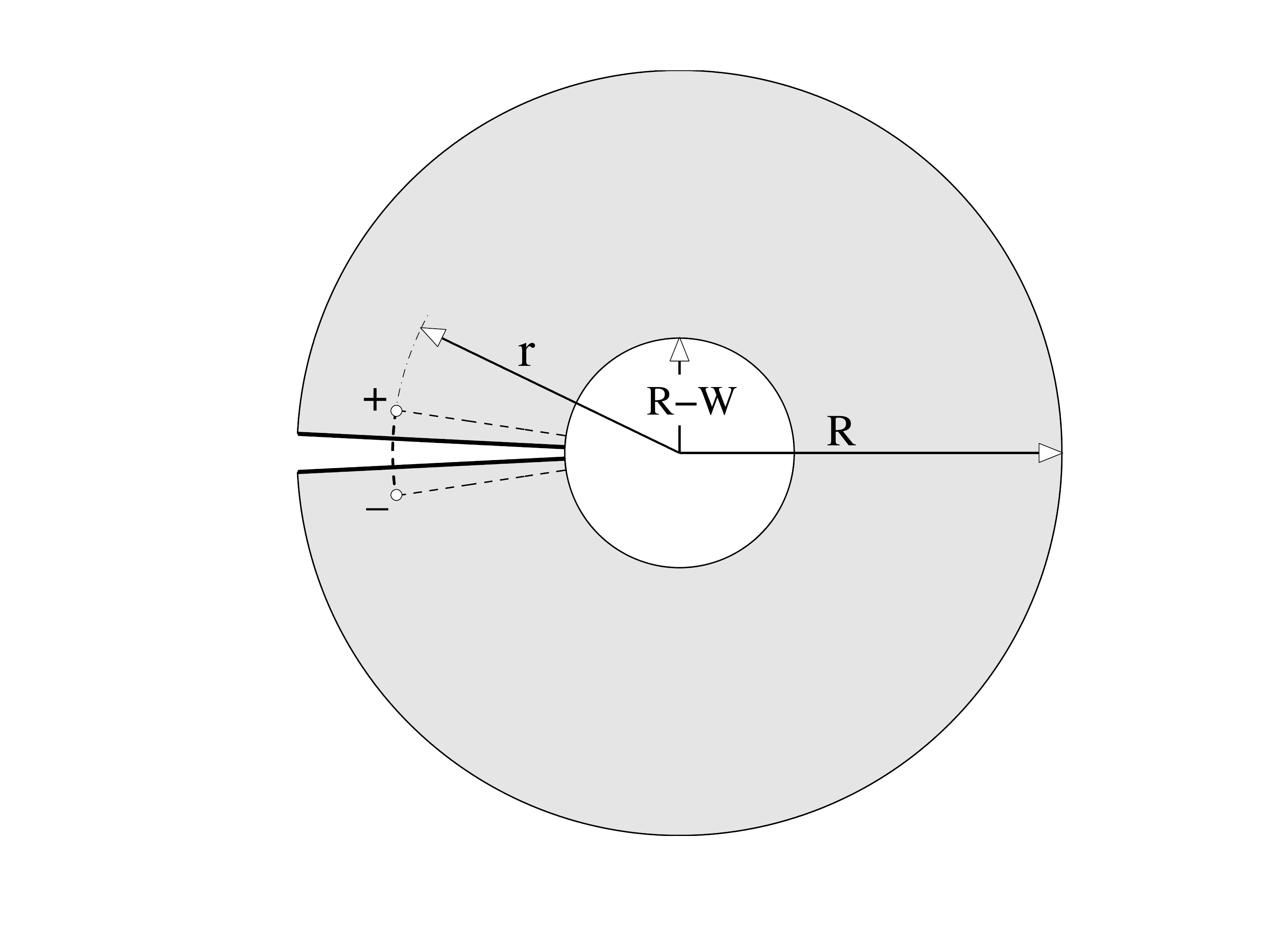}}
\caption[1]{\label{Fig_1a}
The superconducting annulus of radius $R$ and width $W$ with an edge junction to the left. Ideally, the gap between the upper $(+)$ and lower $(-)$ banks tends to zero, but is oversized for visual purposes. Also sketched is the line contour for integration of the phase variations around the junction.}
\end{figure}

The basic methodology used for evaluating the effect of vortices on the response of the junction to the applied magnetic flux was introduced by Clem \textcolor{mycolor}{in the rectangular strip geometry.\cite{ref:clem_2011,ref:clem_2010}} \textcolor{mycolor}{To start with, in view of the conditions reported in the reference experimental work} (Ref.\onlinecite{ref:golodp}) one may consider a quasi-planar system. \textcolor{mycolor}{In our case, a convenient representation of the superconducting loop interrupted by an edge-type junction is realized by an open annulus with infinitesimally small thickness and aperture as sketched in Fig.\ref{Fig_1a}. The physically relevant dimensions of the loop $W$ and $R$ are also defined. Hereafter, the upper and lower limits of the non-superconducting gap are respectively denoted as `` $+$'' and ``$-$'' bank.}

\textcolor{mycolor}{From the physical point of view, a minimal model that allows to predict the $I_c(B)$ pattern should include the electrodynamic field quantities, as well as the superconducting order parameter. A combination of the second London and Ginzburg-Landau equations will do.}
%
%
%
So that, considering that the sample lies on the $XY-$plane, the second London equation modified by the presence of an individual vortex reads
\textcolor{mycolor}{
\begin{equation}
\label{eq:london3D}
\mu_{_0}{\bf curl}(\lambda^2\,  {\bf j})+{\bf B}=\Phi_{_0}\delta({\bf r}-{\bf r}_v)\, \hat{\bf z} \, ,
\end{equation}
}
with ${\bf r}_v$ indicating the position of the vortex, \textcolor{mycolor}{$\delta$ the 2D delta function, and as customary, $\mu_0$ standing for vacuum's permeability and $\Phi_0$ for the flux quantum. $\lambda$ is the London's penetration depth. As it will be assumed that the applied magnetic field is along the $Z-$axis, so will be the full magnetic induction ${\bf B}=B\,\hat{\bf z}$ (also contributed by screening currents and vortices).}

Owing to the quasi-2D nature of the problem, the dependence on the third coordinate may be neglected. \textcolor{mycolor}{Thus, if the film thickness $d$ is much less that $\lambda$, the fields are nearly uniform across the thickness and their spatial variation may be obtained by a quasi-2D approximation.\cite{ref:pearl} It may be obtained as follows. When Eq.(\ref{eq:london3D}) is integrated over the uniform film thickness, say $-d/2<z<d/2$, and in terms of the {\em integrated current density} ${\bf g}~\equiv~\int_{-d/2}^{d/2}~{\bf j}(x,y,z)~dz$} one gets the 2D equation
\begin{equation}
\label{eq:london}
\mu_{_0}\Lambda\, {\rm curl}_2({\bf g})+{B}=\Phi_{_0}\delta({\bf r}-{\bf r}_v) \, .
\end{equation}
Here, we have introduced the {\em effective penetration depth} for thin films $\Lambda=\lambda^2/d$ and \textcolor{mycolor}{${\rm curl}_2({\bf g})$ is the notation for $\partial g_y/\partial x - \partial g_x/\partial y $.}

\textcolor{mycolor}{The property of global charge conservation leads to a useful transformation of the above formula. Thus, in stationary conditions, this is expressed by ${\rm div}({\bf g})=0$, and mathematically, it allows to introduce a scalar quantity, the so-called {\em stream function} $\sigma(x,y)$} such that
\textcolor{mycolor}{
\begin{equation}
\label{eq:stream}
{\bf g}= \frac{\Phi_{_0}}{2\pi\mu_{_0}\Lambda}{\bf curl}(\sigma\,\hat{\bf z}) \, .
\end{equation}
}
\textcolor{mycolor}{London's equation takes the form}
\textcolor{mycolor}{
\begin{equation}
\label{eq:london_mod}
\displaystyle{
\frac{\partial^2 \sigma}{\partial x^2}+\frac{\partial^2 \sigma}{\partial y^2} = \frac{2\pi}{\Phi_0}{B}-2\pi\delta({\bf r}-{\bf r}_v)\, .
}
\end{equation}
}

\textcolor{mycolor}{On the other hand, the counterpart of global charge conservation in the properties of the superconducting condensate is formulated in terms of gauge invariance.\cite{ref:weinberg} Namely, this is implied by the second Ginzburg-Landau equation}
\textcolor{mycolor}{
\begin{equation}
\label{eq:landau}
{\bf g}=-\frac{\Phi_{_0}}{2\pi\mu_{_0}\Lambda}|f|^2\left({\bf grad}(\theta)+\frac{2\pi}{\Phi_{_0}}{\bf A}\right) \, ,
\end{equation}
}
\textcolor{mycolor}{where $|f|$ is the magnitude of the order parameter, whose suppression in the assumed experimental conditions will be neglected  (i.e.: for weak applied field $|f|^2\approx 1$ almost everywhere), $\theta$ stands for the corresponding phase, and ${\bf A}$ is the electromagnetic vector potential. Recall that Eq.(\ref{eq:landau}) implies that gauge transformations for ${\bf A}$ (${\bf A}\mapsto{\bf A}+{\bf grad}\,\chi$) must be consistent with the expression of the order parameter so as to ensure that ${\bf g}$ is a well-defined physical quantity, independent of the selection of gauge for ${\bf A}$. Thus, in general, the evaluation of phase variations (the central quantity for deriving the Josephson's critical current) must be done in terms of the so-called {\em gauge invariant phase difference}. This is defined by integration of the bracketed quantity in the above equation along some path that connects two given points within the superconductor}

\textcolor{mycolor}{
\begin{equation}
\label{eq:gauge_phase}
\Delta\phi = \Delta\theta + \frac{2\pi}{\Phi_{_0}}\int {\bf A} \cdot d{\ensuremath{\boldsymbol\ell}}\, .
\end{equation}
}
\textcolor{mycolor}{Along the article, we will be focused on phase variations across the junction, and thus $\Delta\phi$ identified with the variation between banks, i.e.: $\Delta\phi\equiv \phi^{+}- \phi^{-}$ .}

\textcolor{mycolor}{Interestingly, under specific conditions that will be valid for later purposes, the value of $\Delta\phi$ across the junction may be evaluated without explicit expression of either $\theta (x,y)$ or ${\bf A}(x,y)$. For instance, based on the vanishingly small width of the junction, the gauge invariant phase difference between the neighbouring points highlighted in Fig.\ref{Fig_1a}, may be obtained by integration of Eq.(\ref{eq:landau}). This is done along the indicated closed circuit, that embraces a non-superconducting gap. We recall that the related discontinuity is essential to obtain a nonzero value for the phase difference between such infinitesimally close points. Then, if one assumes that the current density ${\bf g}$ flows antiparallel along the radial branches of the circuit it follows}
\textcolor{mycolor}{
\begin{equation}
\label{eq:gauge_phase_junction}
\Delta\phi (r) = \frac{4\pi\mu_{_{0}}\Lambda}{\Phi_0}\int_{R-W}^{r} g_{\rho}^{+}(\rho)\, d\rho\, .
\end{equation}
} 
\textcolor{mycolor}{Here $g_{\rho}$ is the radial component of the current density vector, and we have used $g_{\rho}^{-}(\rho)=-g_{\rho}^{+}(\rho)$. In practice, it will be useful to write $g_{\rho}$ in terms of the {\em stream function} with help of Eq.(\ref{eq:stream}). The additive constant $\Delta\phi (R-W)$ has been chosen to be zero.}

\textcolor{mycolor}{
Thus, our physical problem (response of the annulus to a uniform field in the presence of vortices) is described by the solution of Eqs.(\ref{eq:london}) and (\ref{eq:landau}) within the superconductor under appropriate boundary conditions. Eventually, (\ref{eq:london}) will be replaced by (\ref{eq:london_mod}), and (\ref{eq:landau}) implemented by means of Eq.(\ref{eq:gauge_phase}) or (\ref{eq:gauge_phase_junction}).
The average of $\Delta\phi (r)$ along the junction banks will lead to the $I_c(B)$ pattern by means of the Josephson's zero-voltage relation, i.e.:}
\textcolor{mycolor}{
\begin{equation}
\label{eq:josephson}
I_J\propto\frac{1}{W}\int_{R-W}^{R} {\rm sin}[\Delta\phi(r)]dr  \, .
\end{equation}
}

Before proceeding to derive the actual solution under different conditions of interest, some comments \textcolor{mycolor}{related to the approximations and methodology that will be used} are due: 

\begin{itemize}[wide=0pt, leftmargin=\dimexpr\labelwidth + 2\labelsep\relax]
\item[(i)] Owing to  linearity, the physical quantities that appear in Eqs.(\ref{eq:london}) and (\ref{eq:landau}) may be calculated by addition of contributions from the \textcolor{mycolor}{applied external source, the London screening currents and by the vortex. For instance, within the sample ${\bf g} = {\bf g}_{_{\rm L}} +{\bf g}_v\; ,\;B = B_a +B_{_{\rm L}} +B_v $ and so on.  When implemented straightforwardly}, this gives way to a complicated integro-differential statement (\textcolor{mycolor}{magnetic fields are integrals of current densities}). However, two important simplifications may be used. Following Ref.\onlinecite{ref:clem_2011}, we will assume that the superconductor's width obeys \textcolor{mycolor}{ the very thin limit relation $W~\ll~\Lambda$. Then, as the self-field of the screening currents is of the order of $\mu_{_0}g$, when compared to the first term of Eq.(\ref{eq:london}) it scales as $W/\Lambda$ and may be neglected, contrary to the applied field $B_a$.\cite{ref:moshe} Then, one may write $B_a+B_{_{\rm L}}\to B_a$. Also, in this conditions a  point-like description of the Pearl vortex is justified. As the screening currents flow over distances scaled by $\Lambda$, the vortex will be well described by its strongly diverging field near the core.\cite{ref:clem_2011,ref:koganb}} In addition, although present, Josephson currents  \textcolor{mycolor}{and related fields} will be considered very small as compared to the screening term. \textcolor{mycolor}{Parametrically, this is expressed through the condition $W\ll\ell$ with $\ell$ the characteristic ``thin-film Josephson length'' $\Phi_0/4\pi\mu_{_0}\Lambda g_c$,\cite{ref:moshe,ref:clem_2011} $g_c$ being the junction's integrated critical current density.}

\item[(ii)] Resorting to complex variable techniques will help us to obtain an exact solution for the above equations in the annular geometry. \textcolor{mycolor}{A dedicated method that will be the basic tool for analyzing the behavior of the loop with (or without) applied magnetic field and/or vortices is briefly described next. In-depth explanations and application details may be found in the appendix.}
\textcolor{mycolor}{Accommodating to the standard nomenclature, from this point onward, the points of the sample's $XY-$plane will be denoted in complex notation: $z=x+iy$.}
\end{itemize}

\section{Solution of the problem: annuli without/with vortices}
\label{sec:solution}
\subsection{Calculation method}
As an outstanding benefit of the complex plane representation, we recall that the construction of a conformal transformation $f(z)$ that maps the region of interest (\textcolor{mycolor}{the superconducting film in our case}) onto the upper half-plane is a \textcolor{mycolor}{convenient and} well developed technique.\cite{ref:nehari} \textcolor{mycolor}{In brief, one solves the ``transformed'' constitutive equations and boundary conditions in the half-plane, and then converts back the solution to the original domain. This procedure has been already used in the investigation of Josephson junction problems for the straight strip geometry.\cite{ref:clem_2011,ref:koganb,ref:koganc,ref:kogana} In such cases, explicit formulations of the vortex states have been derived by taking advantage that Eq.(\ref{eq:london_mod}) remains invariant under conformal transformations when $B\approx 0$. Here, we will exploit this idea for the annular geometry and additionally show that one may also use mapping techniques to obtain the solution under applied magnetic induction ($B\neq 0$).}

For the \textcolor{mycolor}{problem of a Josephson junction in the} superconducting loop (see Fig.\ref{Fig_1a}), the original \textcolor{mycolor}{domain} is the open annulus. \textcolor{mycolor}{A convenient composition of transformations for solving the full physical problem, i.e.: screening currents plus the influence of vortices, is sketched below}

\vspace{.1cm}
\centerline{\rule{.5\textwidth}{0.25pt}}
\vspace{-.5cm}
\textcolor{mycolor}{
\begin{eqnarray}
\label{eq:sketch_confomap}
{\tt PHYSICAL\;PROBLEM}
&\quad
\cdots
\quad&
\begin{array}{c}
{\rm open\;annulus}
\nonumber
\\
\quad z-plane\, (x+iy)
\end{array}
\nonumber
\\
\nonumber
\\
&\qquad\Big\downarrow\quad {\rm log}(z)&
\nonumber
\\
\nonumber
\\
\begin{array}{c}
{\tt Response}
\nonumber
\\
{\tt to\;applied\;field}
\end{array}
&\quad
\cdots
\quad&
\begin{array}{c}
{\rm rectangle}
\nonumber
\\
\quad t-plane\, (r+is)
\end{array}
\nonumber
\\
\nonumber
\\
&\qquad\Big\downarrow\quad {\rm sn}(t)&
\nonumber
\\
\nonumber
\\
{\tt Vortex\;states}
&\quad
\cdots
\quad&
\begin{array}{c}
{\rm half-plane}
\nonumber
\\
\quad w-plane\, (u+iv)
\end{array}
\end{eqnarray}
}
\centerline{\rule{.5\textwidth}{0.25pt}}
\vspace{.1cm}

\textcolor{mycolor}{Thus, the original annulus lies on the $z-$plane. As shown in the appendix, by means of a logarithmic map, $t\sim{\rm log}(z)$, it is converted into a rectangle in the so-defined complex $t-$plane, where one solves for the screening currents. Eventually, the rectangle may be mapped onto the upper half-plane, by means of the so-called {\em Jacobi elliptic function} $w\sim{\rm sn}(t)$, and that is the natural domain for solving the vortex states (see appendix A for detailed expressions and their implementation).}

\subsection{Solution of the vortex-free state}

Let us proceed by first considering the problem of the annulus without vortices, i.e.: we solve the special case of Eq.(\ref{eq:london}) given by  
\begin{equation}
\label{eq:londonw}
\mu_{_0}\Lambda\, {\rm curl}_2({\bf g}_{_{\rm L}})+{B_a}=0 \, ,
\end{equation}
where we have introduced the subindex ``${_{\rm L}}$'' to indicate that \textcolor{mycolor}{the related quantity represents the} response of the superconductor \textcolor{mycolor}{through London screening supercurrents, in the absence of vortices. This is to be distinguished from the eventual full problem (with both contributions and ${\bf g}$ unlabeled). Also, we have used the approximation $B\to B_{a}$ as explained above. The applied magnetic induction ${B_a}$, is assumed to be uniform and perpendicular to the plane}. \textcolor{mycolor}{As said,} this equation may be transformed into a scalar version for the stream function
\begin{equation}
\label{eq:poisson}
\displaystyle{
\frac{\partial^2 \sigma_{_{\rm L}}}{\partial x^2}+\frac{\partial^2 \sigma_{_{\rm L}}}{\partial y^2} = \beta\, ,
}
\end{equation}
where a characteristic scale related to the applied magnetic field and loop's width \textcolor{mycolor}{is defined}: $\beta\equiv 2\pi W^2 B_a /\Phi_{0}$, and distances are assumed to be expressed in units of $W$. Recall that Eq.(\ref{eq:poisson}) is a Poisson-type equation.

\textcolor{mycolor}{Now, the utility of complex variables stems from the fact} that conformal mapping {\em quasi}-preserves Poisson's equation under special circumstances, which apply here. To be specific, when the above equation is transformed to $t-$plane variables, it becomes
\begin{equation}
\displaystyle{
\frac{\partial^2 \varsigma_{_{\rm L}}}{\partial r^2}+\frac{\partial^2 \varsigma_{_{\rm L}}}{\partial s^2} =-\left| \frac{dz}{dt}\right|^2 \beta \, .
}
\end{equation}
as long as the $z(x,y)-$plane is conformally mapped to the $t(r,s)-$plane, \textcolor{mycolor}{i.e.: $t(z)$ is an analytic function. Then, by} replacing $z(t)$ with the \textcolor{mycolor}{inverse} expression of the logarithmic map that transforms the rectangle into the annulus, \textcolor{mycolor}{i.e.: }$z={\rm exp}(-a-it)$ \textcolor{mycolor}{one gets} the statement
\begin{equation}
\label{eq:poisson_sigma}
\displaystyle{
\frac{\partial^2 \varsigma_{_{\rm L}}}{\partial r^2}+\frac{\partial^2 \varsigma_{_{\rm L}}}{\partial s^2} =-e^{2(s-a)}\, \beta \, .
}
\end{equation}
This statement must be solved for Dirichlet boundary conditions, i.e.: \textcolor{mycolor}{$\varsigma_{_{\rm L}}(r,s)=0$ at the boundaries of the rectangle whose dimensions $(a,2b)$ will depend on the actual dimensions of the annulus (see appendix). Physically, this enforces the current flow to be tangential along the boundaries, as may be checked from the definition of $\sigma$.}

\begin{figure}[!]
{\includegraphics[width=0.45\textwidth]{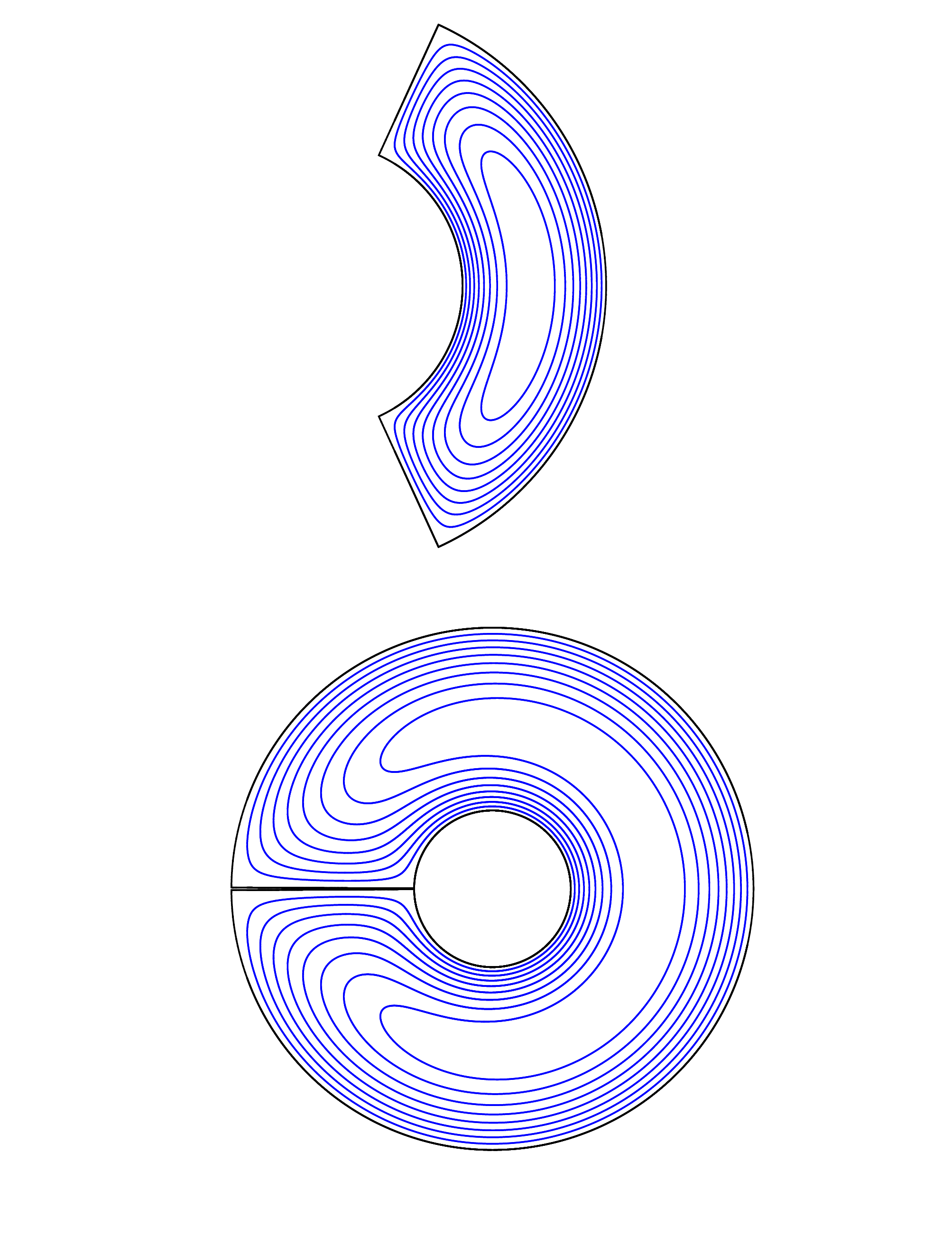}}
\caption[5]{\label{Fig_2}(Color online) 
Contour plot of the stream function $\sigma_{_{\rm L}} (x,y)$ corresponding to a uniform magnetic field induction applied perpendicular to the plane of an open annulus with different apertures.}
\end{figure}

The solution of the above equation within the rectangle \textcolor{mycolor}{is} easily obtained by separation of variables. We get  
\begin{equation}
\label{eq:poisson_varsigma}
\displaystyle{
\varsigma_{_{\rm L}}(r,s)={8\beta}\sum_{n\, odd,m}\, \varsigma_{nm}\, {\rm sin}\left(\frac{n\pi r}{2b}\right){\rm sin}\left(\frac{m\pi s}{a}\right)
}\, ,
\end{equation}
\textcolor{mycolor}{that} is a fast converging series expansion with the coefficients
\begin{equation}
\displaystyle{
\varsigma_{nm}={\rm e}^{-2a}\frac{m[1-{\rm e}^{2a} {\rm cos}(m\pi)]/n}
{(4a^2+\pi^2 m^2)[(n\pi /2b)^2+(m\pi /a)^2]} \, .
}
\end{equation}
The solution within the annulus may be attained by back-substitution,  i.e.
\begin{equation}
\label{eq:sigmah}
\sigma_{_{\rm L}} (x,y)=\varsigma_{_{\rm L}}[r(x,y),s(x,y)]
\end{equation}
Fig.\ref{Fig_2} shows the results obtained from the above series expansion by summing over $10$ values for each index.

As we will see below, when conveniently combined with \textcolor{mycolor}{Eq.(\ref{eq:gauge_phase_junction}), the function $\sigma_{_{\rm L}}(x,y)$} will allow to evaluate the contribution of the screening currents to $\Delta\phi$.

%

\subsection{Vortex states of the loop}

Below, we concentrate on the hereafter called vortex states, that are solutions of \textcolor{mycolor}{London's equation in the form}
\textcolor{mycolor}{
\begin{eqnarray}
\label{eq:vortex}
\mu_{_0}\Lambda\, {\rm curl}_2({\bf g}_{\, v})&=&\Phi_{_0}\delta({\bf r}-{\bf r}_v)
\nonumber\\
&\Downarrow&
\\
\frac{\partial^2 \sigma_{\, v}}{\partial x^2}+\frac{\partial^2 \sigma_{\, v}}{\partial y^2} &=& -2\pi\delta({\bf r}-{\bf r}_v)\, .
\nonumber
\end{eqnarray}
}
Now, the subindex ``$_v$'' indicates that one means to solve for the properties of the loop with zero applied field and a vortex pinned at the position ${\bf r}_v$. \textcolor{mycolor}{We notice that Eq.(\ref{eq:vortex}) leaves out the magnetic induction term $B_v$. As said before, this means that one uses the strong divergence of the Pearl vortex\cite{ref:pearl} at small distances. This may be interpreted as the predominance of the currents' kinetic term of the electromagnetic energy. Correspondingly, concerning the phase variation introduced in Landau's equation, one neglects the vector potential ${\bf A}_v$ as compared to the gradient of the order parameter}

\begin{eqnarray}
\label{eq:landau2}
&\displaystyle{{\bf g}_v=
-\frac{\Phi_{_0}}{2\pi\mu_{_0}\Lambda}{\bf grad}(\phi_{v})
\approx -\frac{\Phi_{_0}}{2\pi\mu_{_0}\Lambda}{\bf grad}(\theta_{v})}
,&\quad ({\bf r}\neq{\bf r}_v)
\nonumber
\\
&\Downarrow& 
\nonumber
\\
&d\phi_v = d\theta_{v} \, .&
\end{eqnarray}
\textcolor{mycolor}{Here, one must recall the requirement for the ``winding number'' of the phase function: $\oint d\phi_v = 2\pi$ if the integration contour embraces the position of one vortex.}

\textcolor{mycolor}{Physically, the functions $\sigma_{v}$ and $\phi_{v}$ fulfilling the above equations represent the Pearl vortex. Mathematically, one may check that $\sigma_{v}(x,y)$ and $\phi_{v}(x,y)$ satisfy Cauchy-Riemann conditions\cite{ref:C_R} and are harmonic for ${\bf r}\neq{\bf r}_v$. This justifies the construction of an analytic function, the complex potential $\psi_{\, v}(z)~=~\sigma_{\, v}(z)~+i\phi_{\, v}(z)$ as a consistent representation. It has been noticed by a number of authors that the logarithmic function in the complex plane is the basis to construct the solution searched.\cite{ref:clem_2011,ref:koganb,ref:koganc} In fact, it presents a divergence in its real part (as required to reproduce the singularity of the vortex) and the $2\pi$ multiplicity in the imaginary part.}

\textcolor{mycolor}{Thus, one may easily find $\psi_{\, v}(w)$ in the image half-plane and transform back to obtain $\psi_{\, v}(z)$ for the actual annulus (see appendix for details). In order to attain the expression of $\psi_{\, v}(w)$ it is useful to recall that our statement is equivalent to finding the complex electrostatic potential for an infinite line of charge, with $\sigma$ equivalent to the electrostatic potential and $\phi$ to the electric flux function. Then, one may just borrow the solution of that case, typically obtained by the method of images. For the case of a vortex at some position $w_{\, v}=u_{\, v}+iv_{\, v}$ in the upper half-plane}, the related complex potential may be expressed\cite{ref:pot}
\begin{equation}
\label{eq:potW}
\Psi(w)={\rm log}\,\frac{w-w_{\, v}}{w-
w_{\, v}^{*}} \, ,
\end{equation}
as deduced from the method of images. \textcolor{mycolor}{$w_{\, v}^{*}$ stands for the complex conjugate of $w_{\, v}$ and represents the position of the ``image vortex''}.

Eq.(\ref{eq:potW}), with $w$ and $w_{\, v}$ replaced by the conformal transformation in Eq.(\ref{eq:map}) leads to the functions $\sigma_{v}(x,y)$ and $\phi_{\, v}(x,y)$ for an individual vortex in our annulus at position $z_{\, v}=x_{v}+iy_{v}$
\begin{eqnarray}
\label{eq:recover}
\sigma_{v}(x,y)&=&{\rm Re}\left\{\Psi[w(z);w_{\, v}(z_{\, v})]\right\}
\nonumber
\\
\\
 \phi_{\, v}(x,y)&=&{\rm Im}\left\{\Psi[w(z);w_{\, v}(z_{\, v})]\right\} \, .
\nonumber
\end{eqnarray}

\begin{figure*}[h]
\centering
\subfigure{
\includegraphics[scale=0.6]{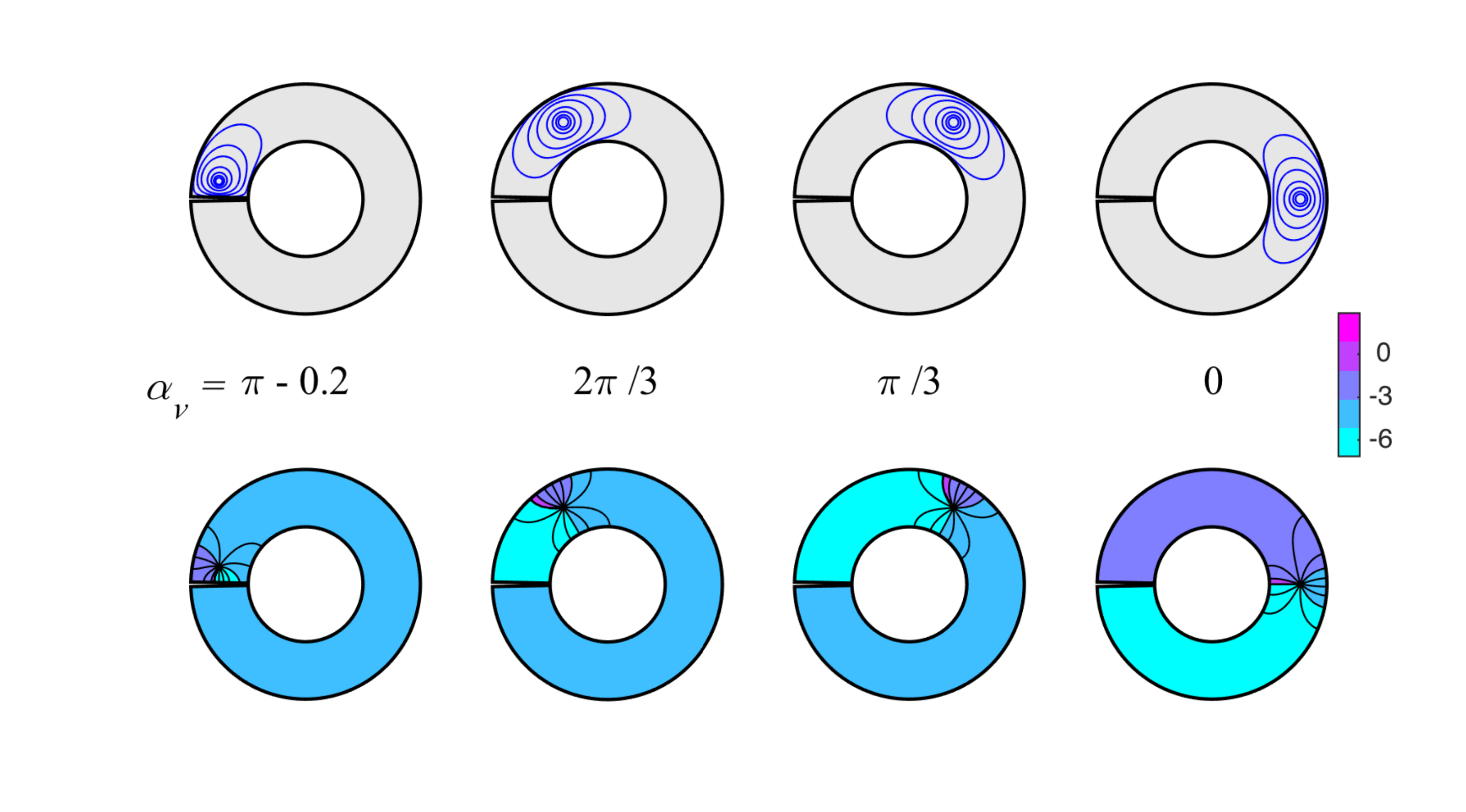}}
\noindent\rule{8cm}{0.4pt}
\subfigure{
\includegraphics[scale=0.6]{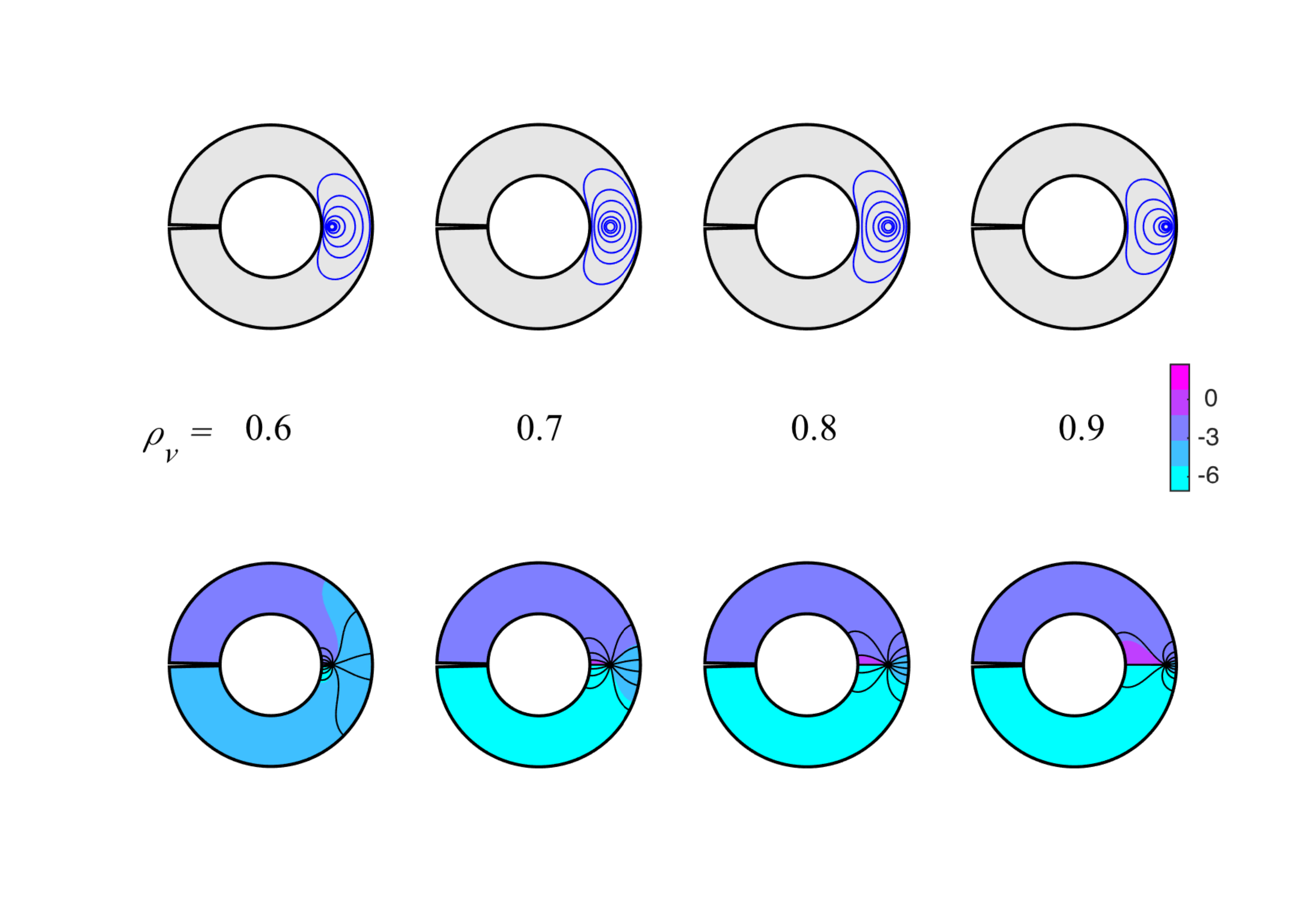}}
\caption{\label{fig:Fig_3} (Color online) Contour plots of the streamfunction $\sigma_{v}(x,y)$ and gauge invariant phase $\phi_{v}(x,y)$ within the superconducting annulus (\textcolor{mycolor}{inner radius} $\rho_{_0}=0.5$) when a vortex is present. Upper panel: the position of the vortex is given by $\rho_{v}=0.75$ and decreasing values of $\alpha_{v}$. For the lower panel: $\alpha_{v}=0$ and increasing $\rho_{v}$. The colormap has been shifted for visual purposes. \textcolor{mycolor}{Radial coordinates are given in units of $R$, i.e.: $\rho=r/R$}.}
\end{figure*}

As an example of what is obtained, Fig.\ref{fig:Fig_3} displays the evolution of the current density streamlines (isolines of $\sigma_{v}$) and the phase $\phi_v$ around a single vortex that settles at given positions within the superconducting loop. Of mention is that noticeable changes in the phase difference between the two banks of the junction may be modulated by trapping the vortex at one point or another. Notice that the $2\pi$ variation of $\phi_{\, v}$ in a closed path around the vortex propagates along the ring, and this may be used to tune the system on demand.

%
%
%
\begin{figure}[th]
\centering
{\includegraphics[width=.52\textwidth]{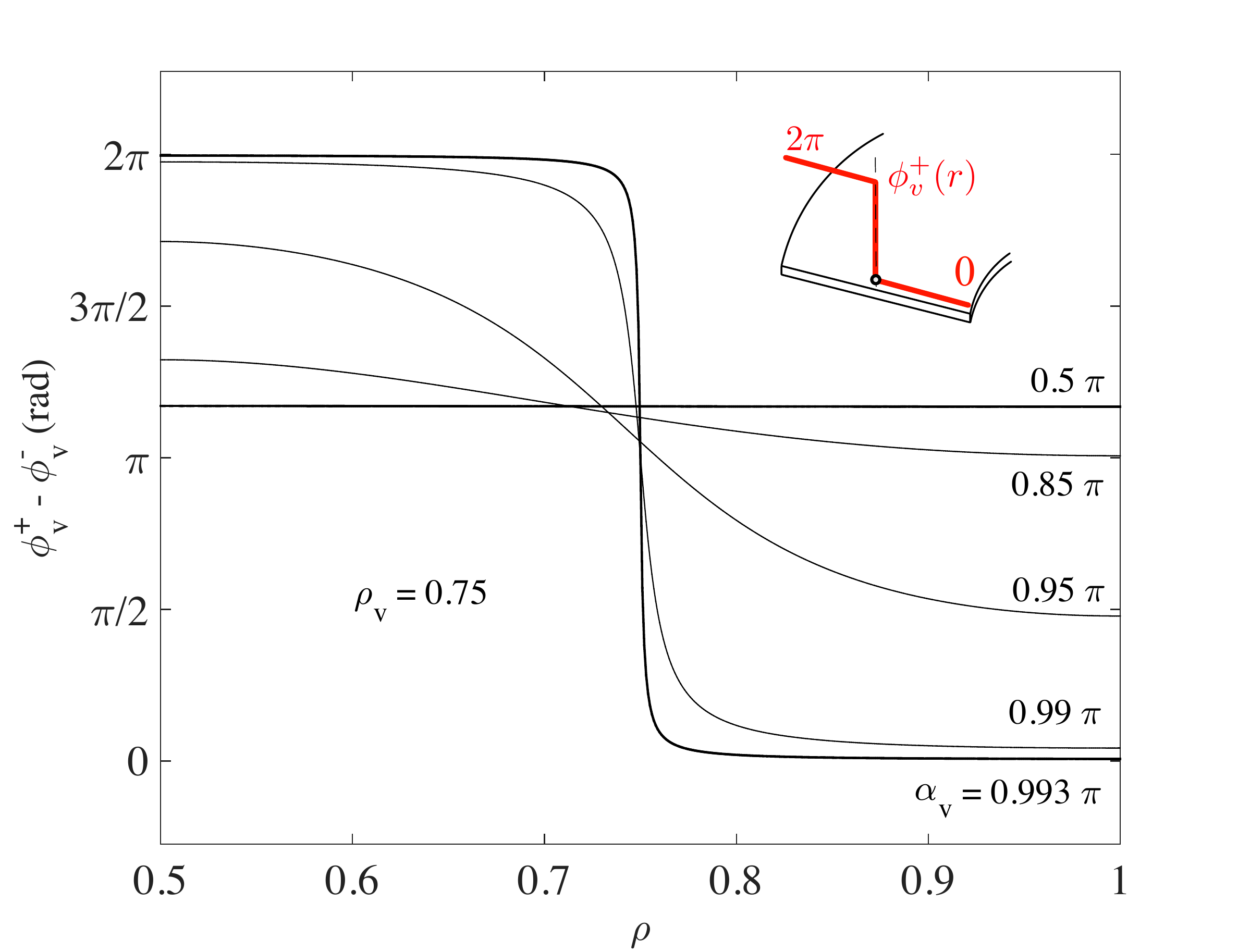}}
\caption[3]{\label{Fig_4}
Gauge invariant phase variation along the banks of the junction for different positions of the vortex given by the coordinates $(\rho_v , \alpha_v)$. \textcolor{mycolor}{As in Fig.\ref{fig:Fig_3}, here} $\rho_{_0}=0.5$. \textcolor{mycolor}{Radial coordinates are relative to the outer radius $\rho\equiv r/R.$} The inset illustrates the limiting situation in which the vortex is very close to the junction.}
\end{figure}
%
\textcolor{mycolor}{
\subsubsection*{Phase variation induced by the vortex, $\Delta\phi_v$}
}
\textcolor{mycolor}{
With the function $\phi_v (x,y)$ at hand, one may calculate variations between any desired couple of points. }
Fig.\ref{Fig_4} offers a detailed view of the phase variation \textcolor{mycolor}{between neighbouring points along the banks of the junction $\Delta\phi_v=\phi_{\, v}^{+}(r)-\phi_{\, v}^{-}(r)$}, induced by a single vortex at different positions in the fashion described in the upper part of Fig.\ref{fig:Fig_3}. As expected,\cite{ref:koganb} when the vortex is very close to the junction \textcolor{mycolor}{($\alpha_v \lesssim\pi$)}, a phase difference of $2\pi$ appears between the inner and outer segments,  \textcolor{mycolor}{as shown in the inset. The reason is that the $2\pi$ variation around the vortex basically occurs in the small separation from this to the closest bank. This fact is visualized in Fig.\ref{fig:Fig_3}. Notice that the contour lines of $\phi_{\, v}$ concentrate in that small region as the vortex gets closer and closer to the junction.} Notice also that, as argued in Ref.\onlinecite{ref:koganb} when the vortex gets farther and farther from the junction, a nearly constant phase difference between the banks occurs (according to Fig.\ref{fig:Fig_3}, in our example this is already valid for $\alpha_v \gtrsim 0.5\,\pi$). Thus, hereafter, although always working with averaged values along the banks, when dealing with distant vortices, we will plainly speak about ``phase difference''.

\begin{figure}[!]
{\includegraphics[width=0.5\textwidth]{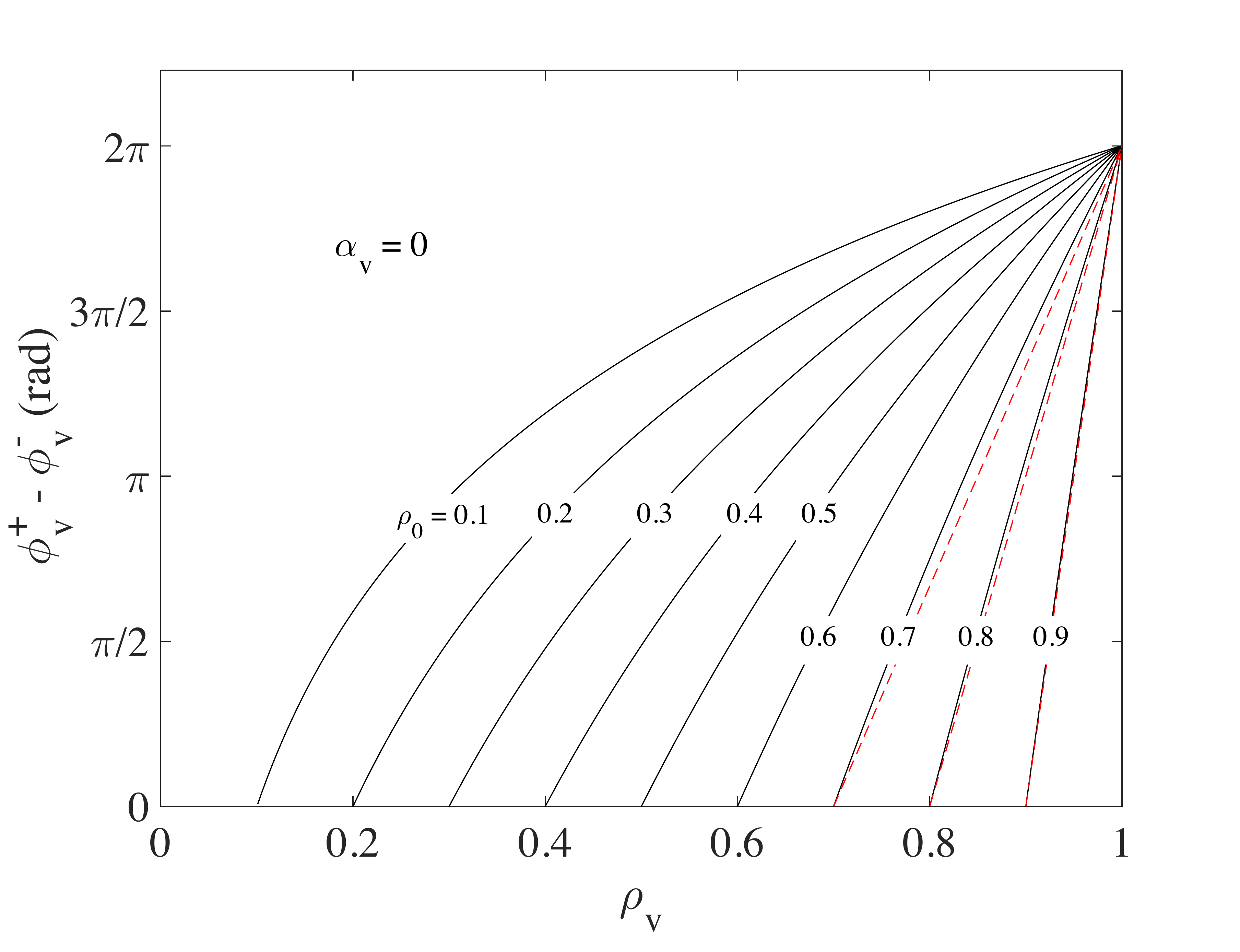}}
\caption[4]{\label{Fig_5}
\textcolor{mycolor}{Averaged phase difference between the two banks of the junction in terms of the radial position of a distant vortex ($\rho_v$) opposite to the aperture $(\alpha_{v}=0)$. The different lines correspond to superconducting loops of different widths given by the value of the inner radius. For each width, the vortex covers the range $\rho_{_0}<\rho_{v}<1$, i.e.: $R-W<r_v<R$). The linear approximation (dashed lines corresponding to Eq.(\ref{eq:linapprox})) is shown for the thinner loops.}}
\end{figure}

The importance of the loop's width to radius ratio is illustrated in Fig.\ref{Fig_5}, that quantifies the effect of a distant vortex on the gauge invariant phase difference. The vortex goes over a set of positions as sketched in the lower part of Fig.\ref{fig:Fig_3}. \textcolor{mycolor}{Now, we plot the averaged (essentially constant) phase difference $\phi_v^{+}-\phi_v^{-}$ between the two banks, as a function of the vortex position, and for different widths of the loop.}

%
%
\textcolor{mycolor}{As an interesting property for the thinner annuli, i.e.: those for which $\rho_0 > 0.6$ or equivalently $W/R<0.4$, a simple dependence of the phase variation in terms of the vortex position is observed (see Fig.\ref{Fig_5}). $\Delta\phi_v$ changes linearly from $0$ to $2\pi$ as the vortex moves from the inner part of the loop ($\rho_v = \rho_{_0}$) to the periphery ($\rho_v = 1 $). Apparently, this fact that may be quantified by using the relation}
\textcolor{mycolor}{
\begin{equation}
\label{eq:linapprox}
\displaystyle{\phi_v^{+}-\phi_v^{-} \approx 2\pi \frac{\rho_v - \rho_0}{1 - \rho_0}}    
\end{equation}
}
\textcolor{mycolor}{We recall that the linear behavior for thin annuli was to be expected, by comparison to such result for long straight loops,\cite{ref:koganb} that can be considered a limiting case of the former.} 

\textcolor{mycolor}{On the other hand, contrary to the expectations, other properties of the semi-infinite strip solution do not straightforwardly extrapolate to this case.\cite{ref:koganb} Thus we find that, when the vortex moves from the inner part of the loop towards the outer part i.e.: $\rho_{_0}<\rho_{v}<1$, the change in phase difference ranges from $0$ to $2\pi$ (Fig.\ref{Fig_5}). It is for a unique intermediate position that one has \textcolor{mycolor}{  $\phi_v^{+}(\rho_v^\pi)-\phi_v^{-}(\rho_v^\pi)=\pi$}. The actual value of the radius $\rho_v^\pi$ for a given geometry may be obtained numerically. As one may deduce from Fig.\ref{Fig_5} this will not generally occur in the midpoint, safe for the thinner loops, for which the linear regime is valid.}

\textcolor{mycolor}{On the other hand, also remarkable is the fact that the $2\pi$ difference between the situations in which the vortex sits either at the inner or outer parts of the annulus, when opposite to the junction (i.e. $\rho_v=\rho_{_0}\;;\;\alpha_{v}=0$ {\em vs.} $\rho_v=1\;;\;\alpha_{v}=0$) occurs independently of the width to radius relation. Once, again this may be understood from the fact that when the vortex is close to some boundary, the overall change of phase from $0$ to $2\pi$ takes place in the small gap in between (see Fig.\ref{fig:Fig_3}).}

\vspace{0.5cm}
\section{Properties of the global phase and the pattern  $I_c(B)$}
\label{sec:globalF}

At this point, we have the elements necessary for the investigation of the influence of nearby vortices in the observable properties of the junction, i.e.: in the critical current dependence $I_c (B)$. As said, \textcolor{mycolor}{the global gauge invariant phase increment $\Delta\phi$ is obtained by addition of the contributions of the superconducting condensate in the absence of vortices $\Delta\phi_{_{\rm L}}$ (with help of Eq.(\ref{eq:gauge_phase_junction})) and the isolated vortex term $\Delta\phi_v$ (with help of Eq.(\ref{eq:recover})).}

\textcolor{mycolor}{For a given annulus, the phase increment along the junction $\Delta\phi (r)$, will be determined by the combined action of the applied magnetic field, implicit in $\Delta\phi_{_{\rm L}}$, and the position of the vortex, implicit in $\Delta\phi_v$. Then, as shown by Clem,\cite{ref:clem_2011} and Kogan and Mints,\cite{ref:koganc} the macroscopic critical current may be obtained by averaging the exponential of the phase difference along the banks. In fact, this statement is equivalent to maximizing the term $\langle {\rm sin}(\Delta\phi)\rangle$ in Josephson's relation.} In normalized units, \textcolor{mycolor}{for the annulus this reads}
\textcolor{mycolor}{
\begin{equation}
\label{eq:maxIc}
\displaystyle{\frac{I_{\rm c}}{I_{\rm c0}} = \frac{1}{1-\rho_{_0}}\left|\,\int_{\rho_{_0}}^{1}e^{i\Delta\phi(\rho)}\, d\rho\,\right|}\, ,
\end{equation}
}
where $I_{\rm c0}$ represents the maximum critical current value, i.e.: corresponding to the phase difference $\Delta\phi= 0$.

\subsection{Contribution of the screening currents}

\begin{figure}[t]
{\includegraphics[width=0.5\textwidth]{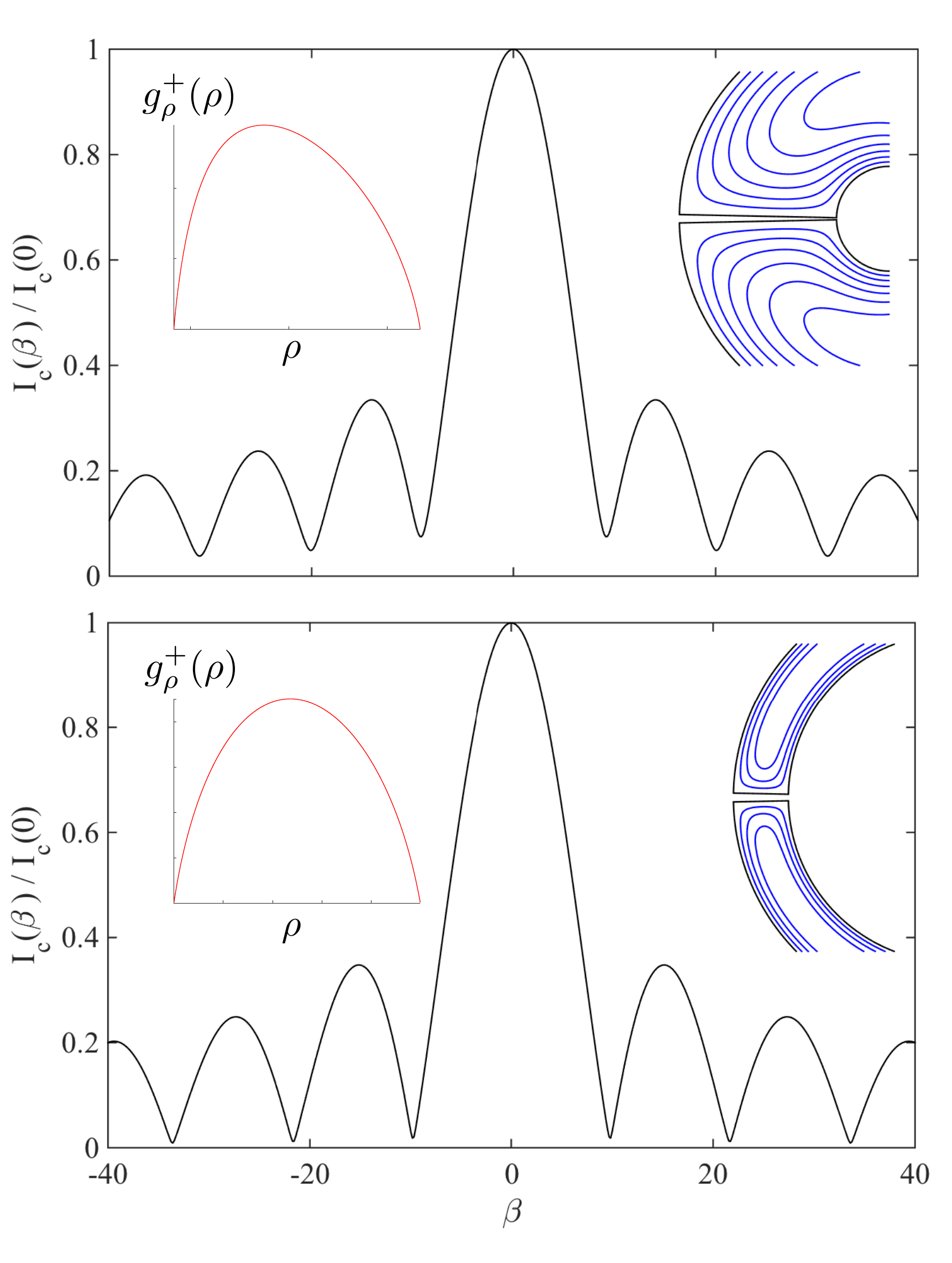}}
\caption[4]{\label{Fig_6}
Critical current pattern in terms of the applied magnetic field with no vortex present. The field is given in dimensionless units $\beta\equiv 2\pi W^2 B_a /\Phi_{_0}$. We plot the results for two different loops with respective normalized widths ${W/R}=0.75$ and ${W/R}=0.25$.
\textcolor{mycolor}{The insets to the right show a detail of the streamlines of the current density ${\bf g}$. To the left, we show the dependence of the radial component $g_{\rho}^{+}$ close to the upper bank of the junction.}}
\end{figure}

Resorting to knowledge on straight strip junctions, and in view of the linear dependence of $\sigma_{_{\rm L}}$ on the applied magnetic field (Eq.(\ref{eq:poisson_varsigma})), one could expect a classical Fraunhofer-like pattern for the critical current dependence $I_c (\beta)$ when no vortices are present in the annulus. However, \textcolor{mycolor}{as displayed in Fig.\ref{Fig_6}}, such property does not hold here. Let us see why. In the former case, \textcolor{mycolor}{say an edge-type junction along the $y-$axis of a long strip parallel to the $x$ direction, $\Delta\phi_{_{\rm L}}(y)$ is well approximated by a sinusoidal dependence.\cite{ref:clem_2010,ref:moshe} 
Then, as $\Delta\phi_{_{\rm L}}$ is an odd function of position respect to the center of the junction, the equivalent to Eq.(\ref{eq:maxIc}) becomes}
\begin{equation}
\displaystyle{\frac{I_{\rm c}(\beta)}{I_{\rm c0}} \sim \left|\frac{1}{\pi}\,\int_{0}^{\pi}{\rm cos}(\beta\, {\rm sin}\tau)d\tau\right|\equiv \left|J_{_0}(\beta)\right|} \, ,
\end{equation}
\textcolor{mycolor}{a familiar Fraunhofer-type diffraction pattern in terms of the normalized magnetic field $\beta$.}

\textcolor{mycolor}{Nevertheless, contrary to that case, for the geometry considered in this article, the phase increment obtained from Eq.(\ref{eq:gauge_phase_junction}) is a non-symmetric function along the junction. When inserted in Eq.(\ref{eq:maxIc}) it leads to the shapes shown in Fig.\ref{Fig_6}, that display an essential difference to the conventional critical current pattern. Perfect destructive interference in the minima does not occur. From the physical point of view, we notice (see Eq.(\ref{eq:gauge_phase_junction})) that $\Delta\phi_{_{\rm L}}(r)$ depends on the radial component of the current density $g^{+}_{\rho}(r)$. As one may verify from the insets of Fig.{\ref{Fig_6}}, asymmetry of this quantity occurs along the junction due to curvature effects. Thus, for  the thicker annuli, current streamlines are more compressed towards the inner region. Consequently, similar to the case of near-field optics,\cite{ref:guillen} the breakdown of the condition of sinusoidal phase difference, leads to such effects as the absence of zeroes in the pattern.
As expected, asymmetry is less and less relevant when the loop becomes more and more narrow, a fact that may be verified in Fig.\ref{Fig_6}: $g_{\rho}^{+}(\rho)$ shows symmetry and minima become practically zeroes when $W/R \lesssim 0.25$.} 

\begin{figure*}[!]
\centering
\subfigure{
\includegraphics[scale=0.6]{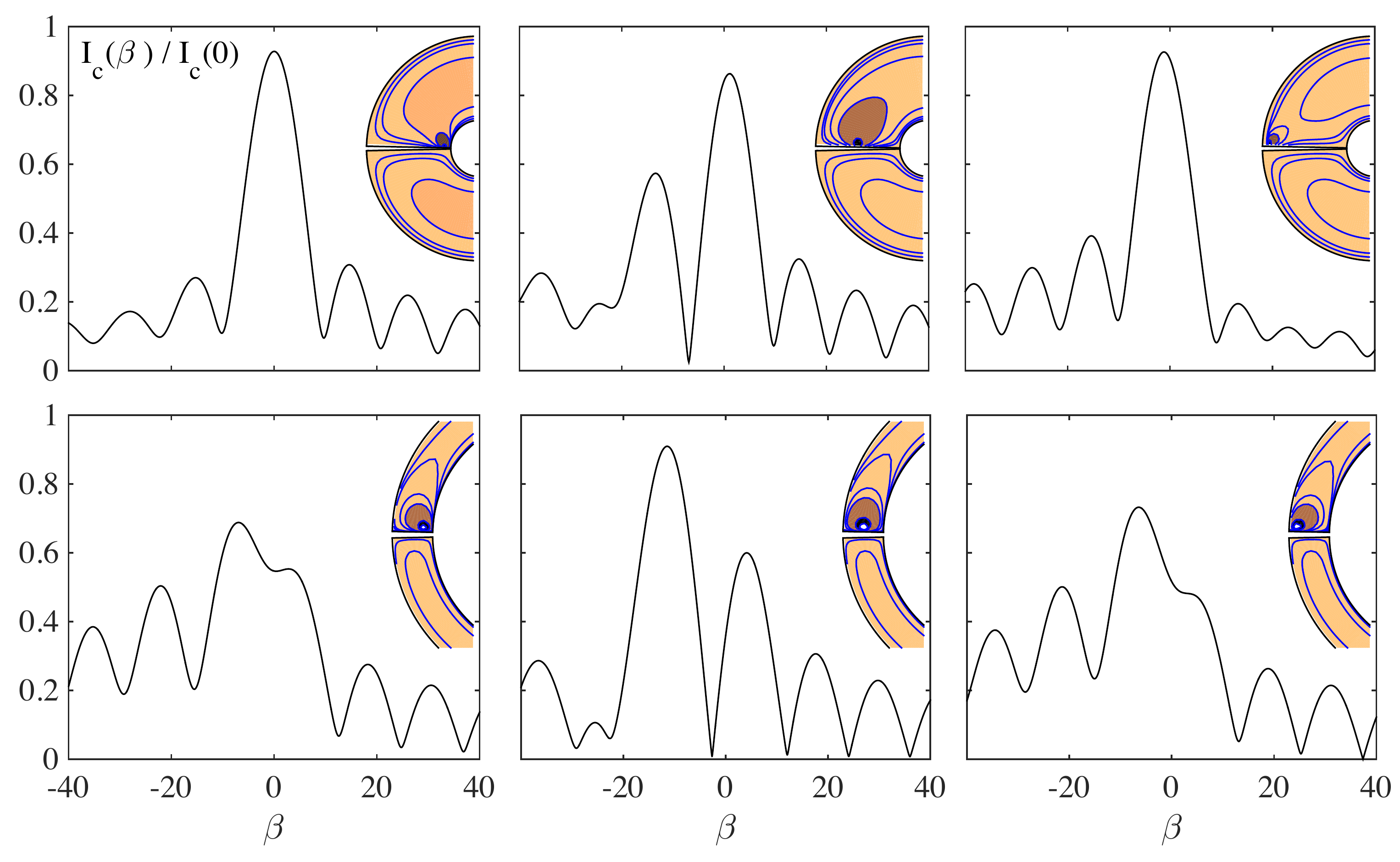}}
\caption{\label{fig:Fig_7} (Color online) Critical current patterns as a function of the applied magnetic field $\beta = 2\pi W^2 B_a /\Phi_{_0}$ in the presence of an antivortex close to the junction in the superconducting loop. The same loops as in Fig.\ref{Fig_6} are considered. The positions of the vortices (revealed by the streamlines $\sigma_{_{\rm L}}+\sigma_v = constant$ in the insets) are $\rho_v = 0.3, 0.625, 0.95$ and $\rho_v =0.8, 0.875, 0.95$ in units of $R$ respectively. \textcolor{mycolor}{Normalized units for $I_c$ are defined in terms of the zero-field, no-vortex value $I_c(0)$.}}
\end{figure*}
%
%

\subsection{Modification of the critical current pattern by the presence of vortices}

Eventually, we analyze the behavior of the critical current pattern $I_c (\beta)$ in the presence of pinned vortices. Eq.(\ref{eq:maxIc}) will be used with $\Delta\phi = \Delta\phi_{_{\rm L}}+\Delta\phi_v$ under different conditions. The basic features are shown in Fig.\ref{fig:Fig_7} for the case of an antivortex at various positions close to the junction. In passing, we comment that the mirror images of the $I_c (\beta)$ patterns respect to the field polarity are obtained by putting a vortex instead. To ease comparison, results for the same loops studied in the previous section are displayed.

As expected from previous literature, the presence of vortices is highly influential,\cite{ref:golodp, ref:golodn, ref:clem_2011, ref:koganb, ref:koganc, ref:mironov} and introduces strong distortions on the vortex-free patterns. Notice that, basically similar to the case of long strip geometries, one finds that the ``middle-position'' response (vortex is equidistant from the boundaries) is characterized by the presence of a minimum instead of a maximum at $\beta$ close to $0$. Remarkably, the value of $I_c$ at the minimum is nearly a zero ($I_c(\beta_{min}) \approx 0$) even for the wide loop. The presence of the vortex or antivortex reverts back to the destructive interference for a specific magnetic field value that depends on the loop's aspect ratio $W/R$.

A relevant distinctive feature of wide loops is that the innermost and outermost positions of the vortices produce clearly unequal  distortions on the $I_c(\beta)$ pattern. Contrary to this, for the narrower loops, one obtains nearly equivalent situations when the vortex is either at the inner or outer radii of the annulus. Again, the behavior of the junction in the narrow loop shares the basic properties of a junction between long straight strips. We recall that in the \textcolor{mycolor}{long} strip limit, a change of phase of $2\pi$ occurs when the vortex position changes from one edge of the strip to the other.\cite{ref:koganb} This leaves the interference pattern unchanged.

\section{Discussion}
\label{sec:discuss}
The \textcolor{mycolor}{circulating} current density and  gauge invariant phase \textcolor{mycolor}{variation} within a superconducting thin film annulus with a Josephson junction have been obtained by combination of the London and Ginzburg-Landau theories. \textcolor{mycolor}{We had a focus on the influence of the loop's geometry and the vortex-induced distortions of the field dependent critical current pattern $I_c(B)$}.

An edge-type junction was considered across the width $W$ of the loop. \textcolor{mycolor}{The exact analytical formulation of the problem has been enabled by a number of suitable approximations. Firstly, assuming the limit of negligible tunneling currents, we have considered an open annulus geometry with a tiny aperture. The simply connected topology of the model simplifies the implementation of conformal mapping techniques to solve the problem, by transforming to rectangular or semi-infinite domains. Secondly, from the physical point of view,} our exact results are valid within the approximation of narrow Josephson junctions (i.e.: $W\ll \Lambda\; ,\; W\ll \ell$ ).\cite{ref:clem_2011} This means that one may \textcolor{mycolor}{unravel} the path for solving the problem by adding the \textcolor{mycolor}{separate} contributions of the screening currents and the vortices. Explicitly we have found an expression for the screening current streamfunction $\sigma_{_{\rm L}} (x,y)$ and for the streamfunction and gauge invariant phase contributed by the vortices $\sigma_v (x,y), \phi_v (x,y)$.

\textcolor{mycolor}{Although the superconducting annulus with an edge-type junction is topologically equivalent to the long strip, they are not equivalent geometrically. This has noticeable consequences on the physical properties, some of which will become unalike as the loop becomes wider and wider.} 

Thus, consistently with the case of long strips, for any value of the ratio $W/R$:
\begin{itemize}
\item We have found that, when the vortex sits at a centered position in between the inner and outer radii, the $I_c(B)$ pattern of the annuli shows a minimum at low fields, instead of a maximum.
\item Noticeable distortions occur as the vortex gets closer to the junction. \textcolor{mycolor}{The consideration of either vortices or antivortices at given positions results in mirror-symmetric profiles of $I_c(B)$}.
\end{itemize}

\textcolor{mycolor}{However,} contrary to the case of long strips, for wide annuli:
\begin{itemize} 
\item Already in the absence of vortices, we find that perfect destructive interference in the $I_c(B)$ pattern no longer occurs. Minima are not zeroes, \textcolor{mycolor}{and this} is more and more noticeable as the ratio $W/R$ increases when ``curvature effects'' are more relevant. This ``node lifting'' effects in our homogeneous superconducting conditions are not to be confused with other predictions or observations related to several kinds of inhomogeneities.\cite{ref:krasnov,ref:neils,ref:lam,ref:kurter,ref:kogana} \textcolor{mycolor}{On the contrary, similar to the case of ``near field'' optics,\cite{ref:guillen} such distortions of the pattern result from asymmetries in the behavior of the phase difference along the junction (as highlighted in Fig.\ref{Fig_6}).}
\item When the vortices get very close either to the inner or outer radius of the loop, one obtains more and more different diffraction patterns as $W/R$ increases. By contrast, in the case of long strips, both situations are completely equivalent (as observed in narrow loops).
\end{itemize}

The methodology introduced in this work can be readily applied to more involved situations as considering the presence of additional vortices or junctions.

\textcolor{mycolor}{Concerning the experimental realization of the effects described,  one may find some possibilities. For instance, a Nb thin film of thickness $d=10$ nm with typical parameters $\lambda \approx 300$ nm and $j_c\approx 10^8$A/m$^2$ would be characterized by $\Lambda\approx 9\,\mu$m, $\ell\approx 14\,\mu$m. This figures leave a reasonable margin to fabricate superconducting annuli with $W$ well below such limits.}

\vspace{0.5cm}

\section*{Acknowledgements}

E. de Lorenzo Poza is gratefully acknowledged for helpful discussions on various mathematical aspects of the problem. 
\textcolor{mycolor}{This work was supported by Spanish Mineco, under projects No. ENE2014-52105-R and No. ENE2017-83669-C4-1-R}.

\appendix
\renewcommand\thefigure{\thesection.\arabic{figure}} 
\setcounter{figure}{0} 
\section{Complex plane representation: conformal mapping}
\label{sec:appendix}

\begin{figure}[h]
\centering
{\includegraphics[width=.45\textwidth]{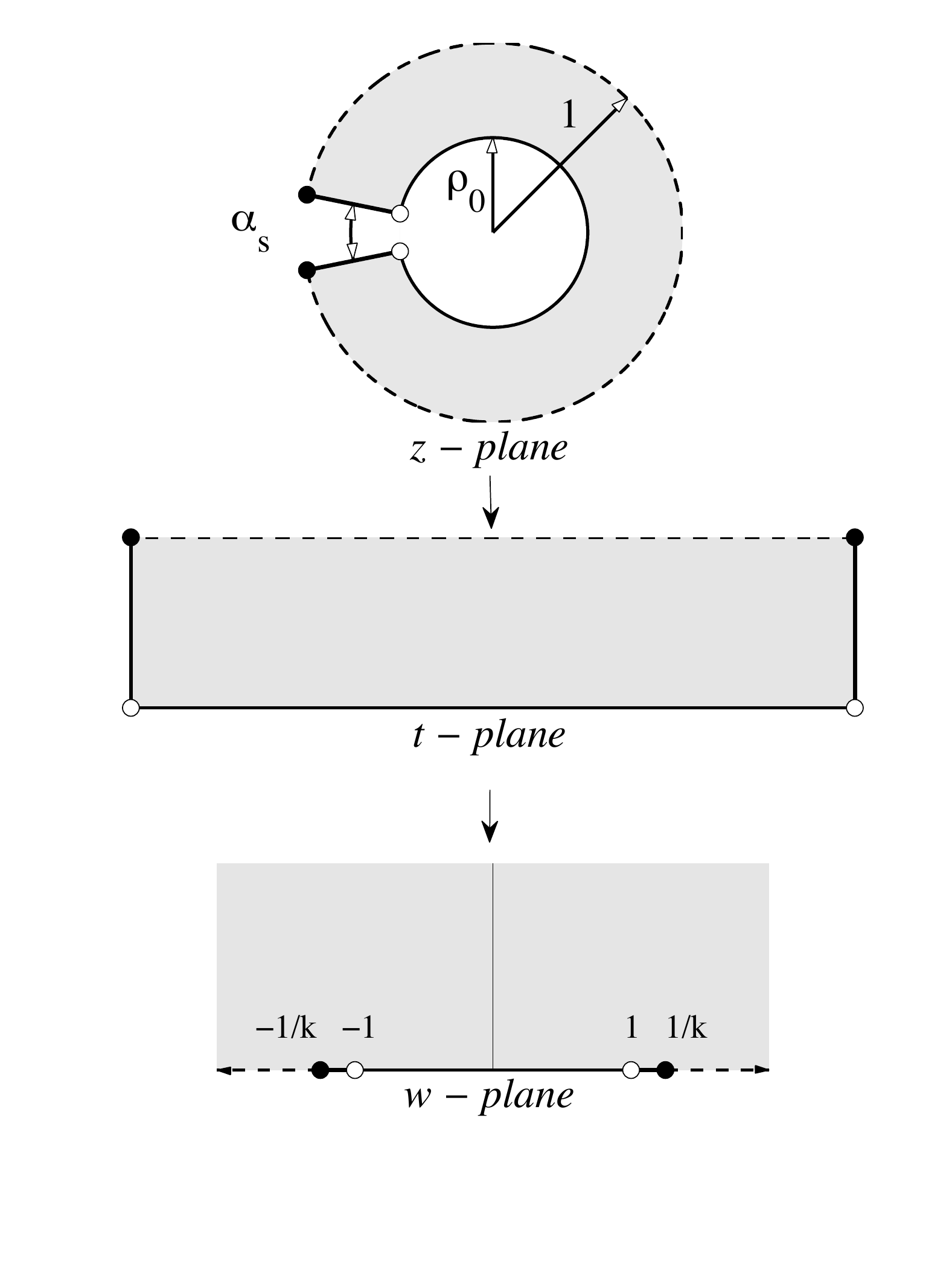}}
\caption[1]{\label{Fig_1}(Color online)
The superconducting loop with a Josephson junction at the left. \textcolor{mycolor}{Schematically shown is the composition of conformal mapping transformations that convert the open annulus into the upper half-plane. Just for visual purposes, the width of the junction is oversized in the picture.}}
\end{figure}

\textcolor{mycolor}{For the case of a Josephson junction in the superconducting loop (see Fig.\ref{Fig_1a}), a possible composition of transformations that facilitates the solution of the physical equations by re-stating in convenient domains is sketched in Fig.\ref{Fig_1}. In terms of the geometrical parameters defined there, the specific conformal mappings that perform such conversion are as follows}

\vspace{0.25cm}
\begin{flushleft}
\boxeqnarray{.475\textwidth}
{
\label{eq:confomap}
\left.
\begin{array}{l}
{\rm open\;annulus\;[\rho_{_0} , \alpha_S]}
\nonumber
\\
\quad z-plane\, (x,y)
\end{array}
\right.
\quad
\xrightarrow{\displaystyle\rm log(z)}
\quad
\begin{array}{l}
{\rm rectangle}\;[2b\times a]
\nonumber
\\
\quad t-plane\, (r,s)
\end{array}
\nonumber
\\
\nonumber
\\
\line(1,0){235}
\nonumber
\\
\nonumber
\\
\left.
\begin{array}{l}
{\rm rectangle\;[2K\times K']}
\nonumber
\\
\quad t-plane\, (r,s)
\end{array}
\right.
\quad
\xrightarrow{\displaystyle\rm sn(t)}
\quad
\begin{array}{l}
{\rm half-plane}\;[v > 0]
\nonumber
\\
\quad w-plane\, (u,v)
\nonumber
\\
\end{array}
\nonumber
}
\end{flushleft}
\vspace{0.25cm}
%
%

In brief, a {\em logarithmic} transformation is used to map the original region (defined by the pair $[\rho_{_0} , \alpha_S]$) into the rectangle $2b\times a \equiv 2\pi-\alpha_S \times {\rm log}(1/\rho_{_0} )$. Secondly, as derived from the Schwarz-Christoffel formula, one maps the rectangle to the upper half-plane\cite{ref:nehari} by means of the parameter dependent {\em Jacobi elliptic function} ${\rm sn}(t|m)$. The actual elliptic function that corresponds to a given rectangle is to be determined through the value of the so-called {modulus} $m$. This is obtained from the double periodicity constraint ${\rm K}'(m)~=~{\rm K}(1-m)$, with ${\rm K}$ the complete elliptic integral of the first kind, and the actual rectangle defined by the relation $2{\rm K}\times{\rm K}' $.

Thus, a proper selection of parameters so as to connect the original region and the upper half-plane reads\cite{ref:jacobian}
\begin{equation}
\label{eq:map}
w = {\rm sn}\left(\frac{iK}{\pi-\alpha_S/2}\,{\rm log}(z)-\frac{iK\,{\rm log}(\rho_{_0})}{\pi-\alpha_S/2}\Big|m\right) \, ,
\end{equation}
where $m$ and the value ${K}\equiv {\rm K}(m)$ are derived by solving the equation
\begin{equation}
\label{eq:Km}
{\rm K}({1-m})+\frac{{\rm log}\,\rho_{_0}}{\pi-\alpha_S/2}\,{\rm K}({m})=0\, .
\end{equation}
We note in passing that the numerical solution of Eq.(\ref{eq:Km}) may be tough for increasing values of the prefactor ${{\rm log}\,\rho_{_0}}/({\pi-\alpha_S/2})$, as machine precision is compromised. This may be a handicap for the investigation of loops with small width ($\rho_{_0}\lesssim 1$). A bypass to this problem has been found through the use of the so-called {\em Nome} special function\cite{ref:abramowitz} $q(m)={\rm exp}(-\pi K'/K)$, whose inverse may be conveniently evaluated in a number of ways. Eventually, the equation to be solved is
\begin{equation}
\displaystyle{m=q^{-1}\left[\rho_{_0}^{1/(1-\alpha_S/2\pi)}\right]} \, .
\end{equation}
From the physical point of view, this technical handicap may be bypassed in practice. In fact, one just needs to use it for verifying that the actual solutions \textcolor{mycolor}{of the equations satisfied by the physical quantities of interest, i.e.: $\sigma (x,y)$ and $\phi (x,y)$}, for narrow loops are well approximated in terms of the results for long rectangles.

\textcolor{mycolor}{Finally, in order to ease discussion about several aspects of the problem, it may be useful to identify the actual image of specific parts of the annulus upon the above transformations. For instance, a vortex close to some boundary of the annulus has an equivalent vortex at some point of the $w-$plane. One may argue in terms of the latter with more ease and revert back general properties of the solution to the real space.}
Thus, as sketched in Fig.\ref{Fig_1}, one may verify that the transformation $w = f(z)$ given by Eq.(\ref{eq:map}) maps the inner circumference of the annulus (radius $\rho_{_0}$) into the segment $[-1,1]$ over real axis, and the banks of the junction to the segments $[-1/k,-1]$ and $[1,1/k]$. The outer circumference (radius $1$) unfolds to cover the rest of the real axis, which acts as the boundary of the superconductor in the upper half of the $w-$plane.

\hfill


\begin{thebibliography}{99}
\bibitem{ref:pot}{It must be clarified that the specific form of Eq.(\ref{eq:potW}) is just a possible selection. It has been assumed that the arbitrary constant determining $\sigma_v$ at the boundary of the loop is $0$. Also, the principal value of the logarithm is used, which determines the arbitrary origin for the phase angle.}
\bibitem{ref:jacobian} {${\rm sn}(z| m)$ stands for the Jacobi elliptic function of complex argument $z$ and modulus $m$. Recall, the less frequent, but not rare use of $k\equiv \sqrt{m}$ in the notation.}
\end{thebibliography}
\end{document}